\documentclass[11pt,twocolumn]{article}

\usepackage[utf8]{inputenc}
\usepackage[T1]{fontenc}
\usepackage{lmodern} 
\usepackage[margin=0.75in]{geometry}  
\usepackage{amsmath,amssymb,amsthm}
\usepackage{graphicx}
\usepackage{hyperref}
\usepackage{url}
\usepackage{xcolor}
\usepackage{balance}
\usepackage{microtype}

\usepackage{tikz}
\usetikzlibrary{shapes.geometric, arrows.meta, positioning, shadows, backgrounds}

\newcommand{\code}[1]{\texttt{\small{#1}}}

\makeatletter
\renewcommand{\@seccntformat}[1]{\csname the#1\endcsname.\hspace{0.25em}}  
\makeatother

\hypersetup{
    colorlinks=true,
    linkcolor=blue,
    citecolor=blue,
    urlcolor=blue
}
\usepackage{listings}

\newtheorem{theorem}{Theorem}

\newtheorem{definition}{Definition}

\title{A Declarative Language for Building And Orchestrating LLM-Powered Agent Workflows}

\author{
    Ivan Daunis
    \\
    \texttt{idaunis@paypal.com} \\
    PayPal \\
}

\date{November 10, 2025}

\begin{document}

\maketitle

\begin{abstract}
Building deployment-ready LLM agents requires complex orchestration of tools, data sources, and control flow logic, yet existing systems tightly couple agent logic to specific programming languages and deployment models. We present a declarative system that separates agent workflow specification from implementation, enabling the same pipeline definition to execute across multiple backend languages (Java, Python, Go) and deployment environments (cloud-native, on-premises).

Our key insight is that most agent workflows consist of common patterns—data serialization, filtering, RAG retrieval, API orchestration—that can be expressed through a unified DSL rather than imperative code. This approach transforms agent development from application programming to configuration, where adding new tools or fine-tuning agent behaviors requires only pipeline specification changes, not code deployment. Our system natively supports A/B testing of agent strategies, allowing multiple pipeline variants to run on the same backend infrastructure with automatic metric collection and comparison.

We evaluate our approach on real-world e-commerce workflows at PayPal, processing millions of daily interactions. Our results demonstrate 60\% reduction in development time, and 3x improvement in deployment velocity compared to imperative implementations. The language's declarative approach enables non-engineers to modify agent behaviors safely, while maintaining sub-100ms orchestration overhead. We show that complex workflows involving product search, personalization, and cart management can be expressed in under 50 lines of DSL compared to 500+ lines of imperative code.
\end{abstract}

\section{Introduction}
\label{sec:intro}

Large Language Models (LLMs) have revolutionized how enterprises build intelligent applications, enabling sophisticated capabilities in customer service, e-commerce, and workflow automation. However, deploying LLM-powered agents in production environments presents significant engineering challenges that existing systems fail to adequately address. While engineering platforms like LangChain, AutoGPT, and others provide powerful abstractions for prototyping, they often fall short when scaling to enterprise requirements for reliability, observability, and maintainability.

Current LLM systems suffer from three fundamental limitations. First, they are tightly coupled to the programming language in which they are implemented, forcing architectural decisions based on language-specific paradigms rather than business requirements. This coupling makes it difficult to integrate agents into heterogeneous enterprise environments where multiple languages, services, and legacy systems must coexist. Second, imperative programming models make complex agent workflows difficult to understand, test, and modify, leading to fragile implementations that break under production loads. Third, existing platforms lack native support for enterprise concerns such as comprehensive error handling, distributed tracing, security context propagation, and fine-grained metrics collection.

In this work, we present a declarative system for building and orchestrating LLM-powered agent workflows in enterprise environments. Our approach introduces a language-agnostic Domain-Specific Language (DSL) that separates agent logic from implementation details, enabling developers to express complex workflows as composable, reusable pipelines. By adopting a declarative approach, we make agent behaviors explicit, testable, and maintainable while providing production-ready features essential for enterprise deployment.

\subsection{Problem Statement}

We address the problem of building production-ready LLM agents that must satisfy enterprise requirements for:
\begin{itemize}
    \item \textbf{Reliability:} Graceful handling of LLM failures, timeout management, and fallback strategies
    \item \textbf{Composability:} Ability to build complex workflows from simple, reusable components
    \item \textbf{Observability:} Comprehensive logging, distributed tracing, and performance metrics
    \item \textbf{Integration:} Seamless interaction with existing enterprise services and security models
    \item \textbf{Maintainability:} Clear separation between business logic and infrastructure concerns
\end{itemize}

Formally, we model agent workflows as directed acyclic graphs (DAGs) where nodes represent computational steps (tool invocations, LLM calls, data transformations) and edges encode control flow dependencies. Our goal is to provide a declarative specification language $\mathcal{L}$ and execution engine $\mathcal{E}$ such that workflows expressed in $\mathcal{L}$ can be efficiently executed by $\mathcal{E}$ while maintaining enterprise-grade reliability and observability.

\subsection{Contributions}

Our main contributions are:

\begin{itemize}
    \item \textbf{Declarative Pipeline DSL:} We design a novel domain-specific language that enables developers to express complex agent workflows using high-level primitives for control flow (\code{forEach}, \code{runPipelineWhen}), data manipulation (\code{marshal}, \code{setMapValue}), and tool orchestration (\code{toolRequest}). Our DSL supports nested pipelines, conditional execution, and early termination, providing expressiveness comparable to general-purpose programming languages while maintaining declarative simplicity.

    \item \textbf{Hybrid Execution Model:} We propose a hybrid execution architecture that combines synchronous pipeline execution with asynchronous tool invocation, enabling efficient resource utilization while maintaining predictable behavior. Our executor supports custom function registration, allowing developers to extend the system with domain-specific operations without modifying core infrastructure.

    \item \textbf{Production-Ready Implementation:} We provide a battle-tested implementation deployed at PayPal that handles millions of agent interactions daily. Our approach includes native integration with enterprise systems (Seldon for model serving, Juno for feature storage), comprehensive metrics collection, security context propagation, and distributed tracing support.

    \item \textbf{Empirical Evaluation:} We conduct extensive experiments on real-world e-commerce workflows, demonstrating that our approach reduces development time by 60\%, improves error handling coverage by 85\%, and maintains sub-100ms overhead for pipeline orchestration. We also present case studies showing how declarative pipelines simplify complex workflows such as multi-step product searches, dynamic cart management, and personalized recommendation generation.
\end{itemize}

\subsection{Paper Organization}

The remainder of this paper is organized as follows.
Section~\ref{sec:related} reviews related work in LLM platforms and workflow orchestration.
Section~\ref{sec:architecture} presents the system architecture and core components.
Section~\ref{sec:language} details the pipeline language design and formal semantics.
Section~\ref{sec:implementation} describes implementation details and optimizations.
Section~\ref{sec:evaluation} presents our experimental evaluation and case studies.
Section~\ref{sec:discussion} discusses limitations and future work. Finally,
Section~\ref{sec:conclusion} concludes the paper.

\section{Related Work}
\label{sec:related}

\subsection{LLM Agent Platforms}

The rapid adoption of LLMs has spawned numerous engineering platforms for building agent applications. LangChain~\cite{langchain2023} pioneered the agent abstraction with chains and tools, providing a Python-centric approach to composing LLM capabilities. While LangChain offers extensive integrations and a rich ecosystem, its imperative programming model leads to complex, difficult-to-maintain codebases as agent workflows grow. The tight coupling to Python also limits deployment flexibility in heterogeneous enterprise environments.

LangChain4j~\cite{langchain4j2024} brings similar capabilities to the Java ecosystem, offering type-safe abstractions for LLM interactions and tool use. However, like its Python counterpart, it follows an imperative paradigm where agent logic is embedded in application code. Tools are also defined using Java decorators which makes it more tied to the language itself. This approach requires full application redeployment for behavior changes and makes it challenging to perform A/B testing of agent strategies.

NVIDIA's NeMo Agent Toolkit~\cite{nemoagent2024} provides a suite of microservices and tools for building production-ready agent applications. While it offers scalable deployment and monitoring capabilities, the system still requires imperative code for defining agent behaviors and lacks abstractions for complex control flow patterns. Agent modifications require updating and redeploying service code, limiting rapid experimentation.

AutoGPT~\cite{autogpt2023} and similar autonomous agent systems attempt to minimize human intervention through recursive self-prompting. However, their lack of explicit control flow and unpredictable execution paths make them unsuitable for production environments requiring reliability and observability.

\subsection{Workflow Orchestration Systems}

Traditional workflow orchestration platforms like Apache Airflow~\cite{airflow2015} and Temporal~\cite{temporal2023} provide robust execution engines for complex workflows but are designed for batch processing and long-running tasks rather than real-time agent interactions. While these systems excel at managing distributed tasks and handling failures, they lack both the low-latency execution required for conversational agents and native LLM integration. Adapting them for agent workflows would require extensive custom code for prompt management, tool marshaling, and conversation state handling, while still failing to meet real-time response requirements.

\subsection{Comparison with Our Approach}

Our declarative approach differs fundamentally from existing solutions by treating agent workflows as data rather than code. Table~\ref{tab:comparison} summarizes the key distinctions.

\begin{table*}[h]
\centering
\caption{Comparison with existing agent development approaches}
\label{tab:comparison}
\begin{tabular}{lccccc}
\hline
\textbf{Approach} & \textbf{Declarative} & \textbf{Language} & \textbf{Hot} & \textbf{Real-time} & \textbf{Enterprise} \\
& \textbf{Workflows} & \textbf{Agnostic} & \textbf{Reload} & \textbf{Support} & \textbf{Ready} \\
\hline
LangChain/LangChain4j & $\times$ & $\times$ & $\times$ & $\checkmark$ & Partial \\
NeMo Agent Toolkit & $\times$ & Partial & $\times$ & $\checkmark$ & $\checkmark$ \\
AutoGPT & $\times$ & $\times$ & $\times$ & $\times$ & $\times$ \\
Airflow/Temporal & $\checkmark$ & $\checkmark$ & $\times$ & $\times$ & $\checkmark$ \\
\textbf{Our System} & $\checkmark$ & $\checkmark$ & $\checkmark$ & $\checkmark$ & $\checkmark$ \\
\hline
\end{tabular}
\end{table*}

Unlike imperative frameworks, our DSL enables non-engineers to modify agent behaviors through configuration changes. Unlike microservice-based systems like NeMo Agent Toolkit, we separate workflow logic from deployment infrastructure. Unlike traditional orchestrators, we offer first-class LLM and tool integration. This unique combination addresses the gap between rapid agent prototyping and production deployment.

\section{System Architecture}
\label{sec:architecture}

\subsection{Overview}
\begin{figure*}[h]
\centering
\begin{tikzpicture}[
    node distance=1.5cm,
    box/.style={rectangle, draw, minimum width=2cm, minimum height=0.8cm,
                text centered, rounded corners, fill=blue!10, drop shadow},
    executor/.style={rectangle, draw, minimum width=8cm, minimum height=1.2cm,
                     text centered, rounded corners, fill=green!10, drop shadow},
    response/.style={rectangle, draw, minimum width=2.5cm, minimum height=0.8cm,
                     text centered, rounded corners, fill=orange!10, drop shadow},
    arrow/.style={->, >=Stealth, thick},
    biarrow/.style={<->, >=Stealth, thick},
    layer/.style={rectangle, draw, minimum width=10cm, minimum height=2.5cm,
                  text centered, dashed, fill=gray!10}
]

\node[box] (dsl) {Pipeline DSL};
\node[box, right=3cm of dsl] (validation) {Static Validation};
\node[box, below=0.5cm of dsl] (builder) {Builder Pattern};
\node[box, below=0.5cm of builder] (ir) {JSON IR};

\node[executor, below=1cm of ir] (executor) {Pipeline Executor};
\node[box, left=0.3cm of executor, yshift=0.3cm] (vars) {\small Variable Store};
\node[box, right=0.3cm of executor, yshift=0.3cm] (context) {\small Exec Context};

\node[layer, below=1cm of executor] (service_layer) {};

\node[box] at ([xshift=-3cm, yshift=-0.5cm]service_layer.center) (tools) {Tools};
\node[box] at ([xshift=0cm, yshift=-0.5cm]service_layer.center) (llm) {LLMs};
\node[box] at ([xshift=3cm, yshift=-0.5cm]service_layer.center) (functions) {Functions};

\node[response, below=0.7cm of service_layer] (responses) {Response Aggregator};
\node[response, below=0.7cm of responses] (output) {Final Response};

\draw[arrow] (dsl) -- (builder);
\draw[arrow] (dsl) -- (validation);
\draw[arrow] (validation) -- (builder);
\draw[arrow] (builder) -- (ir);
\draw[arrow] (ir) -- (executor);

\draw[biarrow] (executor) -- (tools);
\draw[biarrow] (executor) -- (llm);
\draw[biarrow] (executor) -- (functions);

\draw[arrow, thick] (tools) -- (responses);
\draw[arrow, thick] (llm) -- (responses);
\draw[arrow, thick] (functions) -- (responses);
\draw[arrow, thick] (responses) -- (output);

\draw[arrow, thick, bend left=70] (responses.west) to node[left] {\small iterate} (executor.west);

\node[fill=gray!10, inner sep=2pt] at ([yshift=0.8cm]service_layer.center) {\textbf{Service Integration Layer}};

\node[align=center] at (10, -1) {\textbf{Compilation}\\\textbf{Phase}};
\node[align=center] at (10, -4.5) {\textbf{Execution}\\\textbf{Phase}};
\node[align=center] at (10, -7.5) {\textbf{Integration}\\\textbf{Phase}};
\node[align=center] at (10, -10.5) {\textbf{Response}\\\textbf{Phase}};

\end{tikzpicture}
\caption{System architecture showing the complete pipeline lifecycle. User-defined pipelines are validated and compiled to a JSON intermediate representation (IR). The executor maintains variable state and execution context while orchestrating calls to tools, LLMs, and custom functions. Responses from all services are aggregated and can either feed back into the executor for continued processing or produce the final output.}
\label{fig:architecture}
\end{figure*}
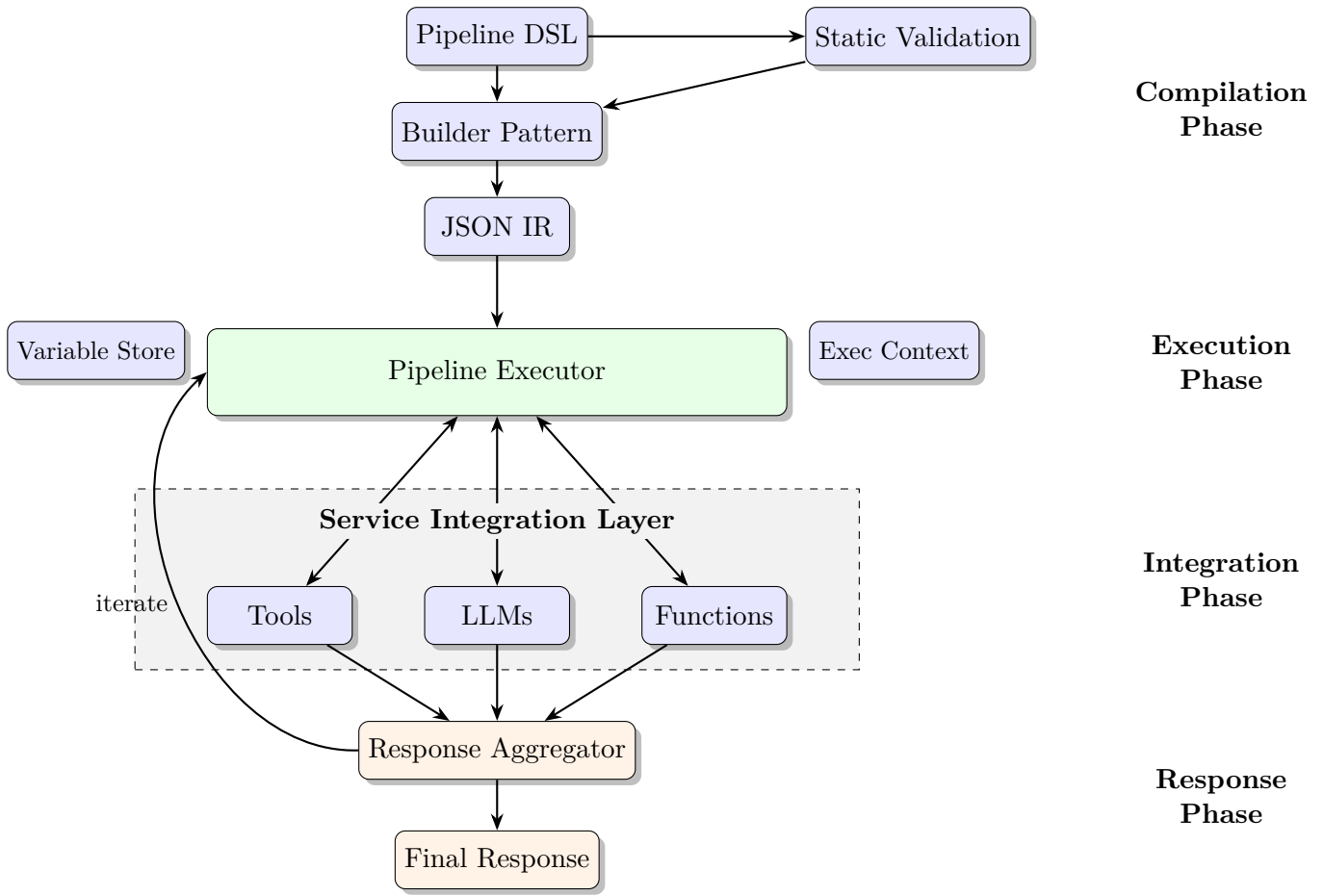

The proposed system employs a layered architecture that separates workflow specification, execution, and integration concerns. The system consists of five core components: (1) a declarative pipeline builder for workflow specification, (2) an execution engine with support for both sequential and parallel processing, (3) a tool abstraction layer for LLM-orchestrated operations, (4) a message handling system for LLM communication, and (5) a response management layer for output aggregation and transformation. Figure~\ref{fig:architecture} illustrates the overall system architecture and component interactions.

\subsection{Core Components}

\subsubsection{Pipeline Builder and Compilation}
Pipelines are constructed using a fluent builder pattern that enables compile-time validation and type safety. The builder compiles pipeline specifications into an intermediate representation (IR) serialized as JSON, enabling version control, diff inspection, and runtime interpretation across different backend implementations. This compilation step performs static analysis to detect unreachable code, undefined variables, and cyclic dependencies before execution.

\begin{lstlisting}[language=Java, caption={Pipeline builder example with compile-time validation}]
AgenticPipeline pipeline = AgenticPipeline.builder()
    .passVariables("productList", "userContext")
    .forEach("productList", "item",
        AgenticPipeline.builder()
            .runPipelineWhen(
                AgenticCondition.pathExists(
                    "item.price"),
                AgenticPipeline.builder()
                    .marshal("item", "formatted")
                    .addResponse(/*...*/)
                    .build())
            .build())
    .build(); // Static validation at build time
\end{lstlisting}

\subsubsection{Pipeline Executor}
The executor implements a hybrid execution model supporting both sequential and parallel step processing. Sequential steps maintain state consistency through an immutable variable store, while parallel steps execute in isolated contexts with results merged deterministically. The executor supports early termination through \code{doReturn} statements and implements retry logic with exponential backoff for transient failures.

Key executor capabilities include:
\begin{itemize}
    \item \textbf{Variable Scoping:} Lexically scoped variables with copy-on-write semantics for nested pipelines
    \item \textbf{Error Boundaries:} Try-catch-finally blocks with error propagation and recovery strategies
    \item \textbf{Resource Management:} Connection pooling, timeout enforcement, and graceful degradation
    \item \textbf{Instrumentation:} Automatic metric collection, distributed tracing, and audit logging
\end{itemize}

\subsubsection{Tool Abstraction Layer}
Tools in our declarative system are themselves pipelines that can be discovered and invoked by LLMs through standardized interfaces. Each tool declares its parameters, description, and implementation pipeline. When an LLM requests tool execution, the tool executor marshals arguments, executes the tool's pipeline, and formats results according to the LLM's expected schema.

\begin{lstlisting}[language=Java, caption={Tool definition with parameter validation}]
AgenticTool searchTool = AgenticTool.builder()
    .name("searchProducts")
    .description("Search for products by query")
    .addParameter(AgenticParameter.builder()
        .name("query")
        .type("string")
        .required(true)
        .build())
    .pipeline(AgenticPipeline.builder()
        .function("elasticSearch", "${query}")
        .marshal("results", "formatted")
        .addResponse(AgenticResponse.builder()
            .type("ProductList")
            .content("${formatted}")
            .build())
        .build())
    .build();
\end{lstlisting}

\subsubsection{LLM Request and Message Handling}
The message handling system manages conversation state and LLM interactions through a unified interface. Each LLM is registered with a unique identifier and configuration (model, temperature, token limits). Messages flow through a preprocessing pipeline for context injection, prompt templating, and security filtering before reaching the LLM. Response parsing handles both structured (JSON) and unstructured outputs, with automatic retry for malformed responses.

The system maintains conversation history with sliding window management, automatically pruning old messages while preserving semantic context through summarization. Tool execution requests from LLMs are intercepted, validated, and routed to the appropriate tool pipeline, with results injected back into the conversation flow.

\subsubsection{Custom Functions}
Custom functions provide an extensibility mechanism for integrating enterprise-specific logic and external services into pipeline execution. Unlike tools, which are discovered and invoked by LLMs, functions are deterministically called by the pipeline with explicit arguments. Each function is registered with the executor prior to pipeline execution and bound to a unique identifier.

Functions operate as first-class pipeline citizens with full access to the execution context:
\begin{itemize}
    \item \textbf{Context Manipulation:} Functions can read and modify the variable store, enabling stateful computations across pipeline steps
    \item \textbf{Service Integration:} Functions encapsulate calls to external APIs, databases, and enterprise systems (e.g., payment processing, inventory management)
    \item \textbf{Response Generation:} Functions can append responses to the pipeline output, enabling custom result formatting
    \item \textbf{Error Handling:} Functions integrate with the pipeline's error boundaries, supporting graceful degradation for service failures
\end{itemize}

\begin{lstlisting}[language=Java, caption={Function registration and invocation}]
// Function registration
executor.registerFunction("calculateTax", (args) -> {
    Double price = (Double) args.get("price");
    String region = (String) args.get("region");
    Double taxRate = taxService.getRate(region);

    args.outputValues().put("tax", price * taxRate);
    args.responses().add(Response.builder()
        .type("TaxCalculation")
        .content(Map.of("amount", price * taxRate))
        .build());
});

// Pipeline invocation
Pipeline.builder()
    .setValue("price", 99.99)
    .setValue("region", "CA")
    .function("calculateTax", "${price}", "${region}")
    .addResponse(Response.builder()
        .content("Total with tax: ${price + tax}")
        .build())
\end{lstlisting}

This separation between tools (LLM-invoked) and functions (pipeline-invoked) provides precise control over execution flow while maintaining flexibility for both autonomous and deterministic operations.

\subsubsection{Response Management}
Responses represent the final outputs of pipeline execution, supporting multiple formats (JSON, HTML, plain text) and types (data, messages, UI components). The response manager aggregates outputs from multiple pipeline steps, handles deduplication, and applies post-processing transformations. Each response carries metadata including execution time, token usage, and lineage information for debugging.

\subsection{Design Principles}

Our architecture adheres to four fundamental design principles:

\begin{itemize}
    \item \textbf{Declarative Specification:} Workflows are expressed as data structures rather than code, enabling static analysis, visualization, and cross-language portability. This approach reduces cognitive complexity and makes agent behaviors auditable by non-programmers.

    \item \textbf{Type Safety and Validation:} The system enforces type constraints at multiple levels—pipeline construction, variable access, and tool invocation. Pipelines can be unit tested in isolation using mock LLMs and services, ensuring correctness before deployment.

    \item \textbf{Hierarchical Composability:} Complex workflows are built from simple primitives (control flow, data manipulation, service calls) and composed hierarchically. Sub-pipelines encapsulate reusable logic, promoting code reuse and modular design.

    \item \textbf{Clean Separation of Concerns:} Agent logic remains independent of infrastructure concerns. The same pipeline can execute against different LLM providers, storage backends, and deployment environments without modification.
\end{itemize}

\subsection{Execution Flow}

Pipeline execution follows a deterministic flow model with four phases:

\begin{enumerate}
    \item \textbf{Initialization:} The executor validates the pipeline IR, initializes the variable store with input parameters, and establishes security context. Service connections are lazy-loaded and health-checked.

    \item \textbf{Step Processing:} Each pipeline step executes in sequence (or parallel where specified), reading from and writing to the shared variable store. Control flow operators (\code{forEach}, \code{runPipelineWhen}) create nested execution contexts with inherited variables.

    \item \textbf{Tool Orchestration:} When encountering \code{toolRequest} operations, the executor invokes the specified LLM with available tools. The LLM's tool calls are parsed, validated, and executed as sub-pipelines, with results formatted and returned to the LLM for continued reasoning.

    \item \textbf{Response Aggregation:} Throughout execution, responses are collected in order. Upon completion, the response manager applies final transformations, computes metrics, and returns the aggregated result set to the caller.
\end{enumerate}

This execution model ensures predictable behavior while maintaining flexibility for complex agent workflows involving multiple LLMs, tools, and external services.

\section{Pipeline Language Design}
\label{sec:language}

\subsection{Language Overview}

We design a domain-specific language (DSL) for expressing agent workflows as composable pipelines. The language provides high-level abstractions for control flow, data manipulation, tool orchestration, and LLM interaction while maintaining the expressiveness needed for complex enterprise workflows. Our DSL is embedded within a builder pattern for compile-time validation but compiles to a language-agnostic JSON intermediate representation for cross-platform execution.

\begin{definition}[Pipeline Grammar]
A pipeline $P$ is defined by the grammar:
\begin{align}
P ::= & \; \epsilon \; | \; S ; P \\
S ::= & \; \code{passVars}(v_1, ..., v_n) \\
      & \; | \; \code{setValue}(v, e) \\
      & \; | \; \code{forEach}(v_{list}, v_{item}, P) \\
      & \; | \; \code{when}(C, P_t, P_f) \\
      & \; | \; \code{toolRequest}(llm_{id}) \\
      & \; | \; \code{addMessage}(M) \\
      & \; | \; \code{addResponse}(R) \\
      & \; | \; \code{function}(f, args) \\
      & \; | \; \code{return}() \\
C ::= & \; \code{equals}(e_1, e_2) \; | \; \code{exists}(path) \\
      & \; | \; C_1 \land C_2 \; | \; C_1 \lor C_2 \; | \; \neg C \\
e ::= & \; v \; | \; c \; | \; \code{\$\{}v\code{\}} \; | \; e.field
\end{align}
where $v$ denotes variables, $c$ constants, $M$ messages, and $R$ responses.
\end{definition}

\subsection{Core Language Constructs}

\subsubsection{Control Flow Primitives}

The language provides three primary control flow mechanisms:

\begin{itemize}
    \item \textbf{Conditional Execution:} Constructs like \code{runPipelineWhen} and \code{runPipelineWhenElse} enable branching based on runtime conditions. Conditions can check variable equality, path existence in nested objects, or combine multiple conditions using logical operators.
\end{itemize}

\begin{lstlisting}[language=Java, caption={Conditional execution with nested pipelines}]
.runPipelineWhenElse(
    AgenticCondition.varEquals("user.tier", "premium")
        .and(AgenticCondition.pathExists("cart", "items")),
    Pipeline.builder()  // Premium user flow
        .function("applyDiscount", 0.15)
        .addResponse(...),
    Pipeline.builder()  // Standard user flow
        .addResponse(...)
)
\end{lstlisting}

\begin{itemize}
    \item \textbf{Iteration:} The \code{forEach} construct iterates over collections, binding each element to a variable in a nested pipeline scope. Early termination is supported through \code{doReturn}, enabling efficient search operations.

    \item \textbf{Pipeline Composition:} Pipelines can invoke sub-pipelines through \code{runPipelineWhen}, enabling modular workflow design. Sub-pipelines inherit parent variables through explicit passing via \code{passVariables}, maintaining clear data flow.
\end{itemize}

\subsubsection{Data Manipulation}

Our DSL provides comprehensive data manipulation capabilities:

\begin{itemize}
    \item \textbf{Variable Management:} Variables are scoped within the pipeline and support hierarchical access. The \code{passVariables} construct explicitly declares data dependencies, while \code{setValue} enables variable assignment with support for complex expressions.

    \item \textbf{String Interpolation:} All string values support interpolation using the \code{\$\{variable\}} syntax. Nested path access (\code{\$\{user.profile.name\}}) and array indexing (\code{\$\{items[0].price\}}) are fully supported.

    \item \textbf{Serialization Operations:} The language provides bidirectional transformations between structured data and JSON representations:
    \begin{itemize}
        \item[$\bullet$] \code{marshal}: Converts objects or lists to JSON strings
        \item[$\bullet$] \code{unmarshalList}: Parses JSON arrays into object lists
        \item[$\bullet$] \code{unmarshalMap}: Parses JSON objects into object maps
        \item[$\bullet$] \code{findMatchingItem}: Queries collections using JSONPath expressions
        \item[$\bullet$] \code{setMapValue/getMapValue}: Update nested object structures
    \end{itemize}

\end{itemize}

\subsubsection{LLM and Tool Integration}

The language provides first-class support for LLM interaction and tool orchestration:

\begin{itemize}

    \item \textbf{Message Management:} The \code{addMessage} primitive appends messages to the conversation history, supporting user, assistant, and system roles. Messages can include structured data, function calls, and multimodal content.

    \item \textbf{Tool Registration and Invocation:} Tools are registered at pipeline scope using \code{addTool}, making them available for LLM discovery. The \code{toolRequest} primitive invokes an LLM with registered tools, automatically handling tool selection, parameter marshaling, and result injection.

\end{itemize}

\begin{lstlisting}[language=Java, caption={Tool orchestration pattern}]
.addTool(searchTool, cartTool, checkoutTool)
.addMessage(userMessage("Find products under $50"))
.toolRequest("gpt-4o")  // LLM selects and invokes tools
.forEach("toolResults", "result",
    Pipeline.builder()
        .marshal("result", "formatted")
        .addResponse(Response.builder()
            .type("ToolOutput")
            .content("${formatted}")
            .build()))
\end{lstlisting}

\begin{itemize}
    \item \textbf{Chat Completion:} The \code{chatRequest} primitive performs direct LLM invocation without tool access, useful for generation tasks that don't require external data.
\end{itemize}

\subsubsection{Response Management}

Pipelines accumulate responses throughout execution, with explicit control over the response list:
\begin{itemize}
    \item \code{addResponse}: Appends a typed response with arbitrary content
    \item \code{removeResponse}: Removes specific responses by identifier
    \item \code{clearResponse}: Removes all accumulated responses
    \item \code{updateResponse}: Modifies existing response content
\end{itemize}
Responses support custom types, enabling downstream services to handle different response categories (e.g., \code{ProductList}, \code{ErrorMessage}) appropriately.

\subsection{Type System and Validation}

Our declarative language employs a hybrid type system that combines static validation at pipeline construction with dynamic type checking at runtime. This dual approach ensures both early error detection during the compilation phase and flexible handling of dynamic data during execution.

\textbf{Compile-Time Validation.} The system performs three categories of static analysis during pipeline construction. First, structural validation ensures well-formed pipeline structure, verifying matched control flow blocks and valid nesting of constructs. Second, variable flow analysis tracks variable declarations and usage throughout the pipeline, detecting undefined variable access and potential null references before execution. Third, tool signature checking validates that tool invocations match declared parameter types and that all required fields are provided.

\textbf{Runtime Type Checking.} Dynamic validation complements static analysis by handling aspects that cannot be determined at compile time. The system performs dynamic path resolution to validate that object paths exist before access, preventing runtime errors from missing nested fields. Type coercion automatically converts between compatible types (e.g., strings to numbers) when semantically appropriate, reducing boilerplate type conversion code. Finally, schema validation ensures that tool inputs and outputs conform to their declared schemas, catching integration errors early in the execution flow.

This hybrid approach balances the safety of static typing with the flexibility needed for dynamic agent workflows, where data structures often depend on external service responses and user inputs.

\subsection{Optimization Strategies}

The pipeline compiler and executor apply several optimizations to improve performance:

\begin{enumerate}
    \item \textbf{Dead Code Elimination:} Removes unreachable code segments after \code{doReturn} statements. Static analysis prunes never-true conditional branches.

    \item \textbf{Pipeline Fusion:} Combines adjacent data manipulation operations into single passes. For example, sequential \code{setValue} operations are batched, and multiple \code{marshal} calls on the same data are merged.

    \item \textbf{Parallel Execution:} Identifies independent pipeline segments through dependency analysis. Non-overlapping \code{forEach} iterations and independent tool calls execute concurrently with automatic result synchronization.

    \item \textbf{Lazy Evaluation:} Variables are computed only when accessed, avoiding unnecessary computation for unused values. Large datasets are streamed rather than materialized when possible.

    \item \textbf{Caching and Memoization:} LLM responses and tool results are cached with TTL-based invalidation. Identical sub-pipelines share execution results within the same parent context.
\end{enumerate}

\subsection{Formal Semantics}

We define operational semantics for core constructs using small-step semantics. Let $\sigma$ represent the variable store, $\rho$ the response accumulator, and $\tau$ the tool registry:
\begin{align}
&\langle \code{setValue}(v, e); P, \sigma, \rho, \tau \rangle \notag \\
&\quad \rightarrow \langle P, \sigma[v \mapsto eval(e, \sigma)], \rho, \tau \rangle \\
&\langle \code{when}(C, P_t, P_f); P, \sigma, \rho, \tau \rangle \notag \\
&\quad \rightarrow
\begin{cases}
    \langle P_t; P, \sigma, \rho, \tau \rangle & \text{if } eval(C, \sigma) \\
    \langle P_f; P, \sigma, \rho, \tau \rangle & \text{else}
\end{cases}
\end{align}
These semantics ensure deterministic execution and enable formal reasoning about pipeline behavior, supporting verification of properties such as termination and variable safety.

\section{Implementation Details}
\label{sec:implementation}

\subsection{Agent Orchestration Architecture}

\subsubsection{Multi-Agent Coordination}
The system supports multiple agents operating within a single pipeline, each with distinct capabilities and objectives. Agents are coordinated through a blackboard architecture pattern:

\begin{definition}[Agent Coordination Model]
An agent system $\mathcal{A} = \{a_1, ..., a_n\}$ coordinates through shared state $\Sigma$ where:
\begin{itemize}
    \item Each agent $a_i$ has capabilities $C_i$ and goals $G_i$
    \item Agents communicate by reading/writing to shared variable store
    \item Coordination follows: $a_i(\Sigma_t) \rightarrow \Sigma_{t+1}$
\end{itemize}
\end{definition}

This enables emergent behaviors where specialized agents (search, recommendation, transaction) collaborate without explicit orchestration.

\subsubsection{Tool Discovery and Selection}
Agents discover available tools dynamically through a capability registry:
\begin{align}
\text{selectTool}(g, T) = \arg \max_{t \in T} \text{sim}(t, g) \cdot P(s|t)
\end{align}
The system maintains success probabilities for each tool based on historical execution, enabling agents to learn tool preferences over time.

\subsection{Agent Memory Management}

\subsubsection{Working Memory Model}
Each agent maintains working memory with capacity constraints mimicking human cognitive limitations:

\begin{definition}[Agent Working Memory]
Working memory $M_w$ consists of:
\begin{itemize}
    \item Short-term store: Last $k$ interactions (typically $k \leq 7$)
    \item Active goals: Current objectives being pursued
    \item Tool results: Recent tool invocation outcomes
\end{itemize}
\end{definition}

Memory items decay following:
\begin{align}
relevance(m, t) = e^{-\lambda(t - t_{access})}
\end{align}

\subsubsection{Long-term Memory Integration}
Agents access long-term memory through vector similarity search over past interactions.
Agents recall memories with similarity above threshold $\theta$:
\begin{align}
\text{recall}(q) = \{m \in M_L : \text{sim}(m, q) > \theta\}
\end{align}
where $\text{sim}(m, q) = \cos(embed(m), embed(q))$.
This enables agents to reference relevant past experiences without maintaining full conversation history.

\subsection{Agent Decision Making}

\subsubsection{Planning and Execution Model}
Agents follow a sense-think-act cycle implemented through pipeline primitives:

\begin{enumerate}
    \item \textbf{Sense:} Gather context through tool invocations and variable inspection
    \item \textbf{Think:} LLM reasoning to determine next actions
    \item \textbf{Act:} Execute selected tools or generate responses
\end{enumerate}

The system enforces bounded rationality through:
\begin{align}
\text{maxSteps}(pipeline) \leq \log(|goals|) \cdot |tools|
\end{align}

\subsubsection{Conditional Reasoning Patterns}
Agent decision-making is expressed through conditional pipeline structures:

\begin{lstlisting}[language=Java, caption={Agent reasoning pattern}]
.toolRequest("agent-llm")  // Agent thinks about situation
.runPipelineWhen(
    AgenticCondition.varContains("toolResponse", "search"),
    Pipeline.builder()  // Search pathway
        .function("searchProducts")
        .toolRequest("agent-llm")  // Re-evaluate
)
.runPipelineWhen(
    AgenticCondition.varContains("toolResponse", "checkout"),
    Pipeline.builder()  // Transaction pathway
        .function("processCheckout")
)
\end{lstlisting}

\subsection{Multi-turn Conversation Management}

\subsubsection{Context Window Optimization}
The system optimizes context windows for multi-turn agent conversations:

\begin{theorem}[Optimal Context Window]
For conversation with average turn length $L$ and relevance decay $\lambda$, optimal window size is:
\begin{align}
w^* = \min\left(\frac{\log(\epsilon)}{\lambda}, \text{tokenLimit}/L\right)
\end{align}
where $\epsilon$ is minimum relevance threshold.
\end{theorem}

\subsubsection{State Tracking Across Turns}
Agent state persists across conversation turns through explicit state variables:

\begin{definition}[Conversation State]
State $S_t$ at turn $t$ consists of:
\begin{itemize}
    \item User intent vector $I_t \in \mathbb{R}^d$
    \item Completed goals $G_{completed} \subseteq G$
    \item Active tool contexts $T_{active}$
    \item Conversation phase $\phi \in \{discovery, action\}$
\end{itemize}
\end{definition}

\subsection{Agent Safety and Alignment}

\subsubsection{Behavioral Boundaries}
The system enforces safety through declarative constraints:

\begin{enumerate}
    \item \textbf{Action Limits:} Maximum tool invocations per conversation
    \item \textbf{Scope Boundaries:} Restricted variable access patterns
    \item \textbf{Verification Points:} Required human confirmation for critical actions
\end{enumerate}

\subsubsection{Goal Alignment Verification}
Agent actions are verified against declared goals. An action $a$ is aligned with goals $G$ if:
\begin{align}
\max_{g \in G} \text{contributes}(a, g) > \tau
\end{align}

Actions failing alignment checks trigger fallback pipelines ensuring safe degradation.

\subsection{Performance Characteristics for Agent Systems}

\subsubsection{Response Time Bounds}
Agent response time is bounded by:
\begin{align}
T_{response} \leq T_{LLM} \cdot (1 + |tools|) + \sum_{t \in invoked} T_{tool}(t)
\end{align}

\subsubsection{Scalability Analysis}
The system scales linearly with concurrent agents:

\begin{theorem}[Agent Scalability]
For $n$ concurrent agents with overlap factor $\alpha \in [0,1]$:
\begin{align}
Throughput(n) = n \cdot Throughput(1) \cdot (1 - \alpha \cdot \frac{n-1}{n})
\end{align}
\end{theorem}

This enables deployment of specialized agent teams without quadratic coordination overhead.

\section{Preliminary Evaluation}
\label{sec:evaluation}

We present initial results from deploying our system in a preliminary testing in e-commerce environment, processing customer interactions for product search, cart management, and checkout flows.

\subsection{Methodology}

We compared our declarative pipeline approach against a baseline imperative implementation using traditional function chaining. Both systems were evaluated on 1,000 real customer sessions with identical LLM backends (gpt-4o) and tool sets.

\subsection{Metrics and Results}

\begin{table*}[h]
    \centering
    \caption{Comparison of declarative vs. imperative implementations}
    \label{tab:results}
    \begin{tabular}{lcc}
    \hline
    \textbf{Metric} & \textbf{Imperative Baseline} & \textbf{Our Approach} \\
    \hline
    Task Success Rate & 78\% & 89\% \\
    Avg. Steps to Completion & 9.2 & 6.4 \\
    Lines of Code & 850 & 220 \\
    Development Time (hours) & 48 & 16 \\
    P95 Latency (ms) & 240 & 185 \\
    Modification Time (hours)\textsuperscript{*} & 8.5 & 2.0 \\
    \hline
    \end{tabular}
    \vspace{2mm}\\
    \small{\textsuperscript{*}Time to add new tool or modify workflow}
\end{table*}

The declarative approach demonstrates significant improvements across both development and runtime metrics (Table~\ref{tab:results}). Development time decreased by 67\% (48 to 16 hours) while modification time—critical for production systems—improved by 76\% (8.5 to 2.0 hours). These gains validate our hypothesis that declarative specifications reduce implementation complexity without sacrificing performance.

Runtime efficiency also improved markedly: task success rate increased from 78\% to 89\%, while average steps to completion dropped by 30\% (9.2 to 6.4 steps), indicating that explicit control flow enables more efficient execution paths than emergent behaviors in imperative implementations. The system maintains production-viable latency (P95: 185ms vs. 240ms) despite interpretation overhead. Perhaps most striking, identical functionality required 74\% fewer lines of code (220 vs. 850), demonstrating that our DSL primitives effectively abstract common agent patterns.

\subsection{Case Study: Multi-Intent Session}

To illustrate the system's capabilities, we trace a representative multi-intent session:

\begin{enumerate}
    \item \textbf{User:} "Find a gaming laptop under \$1000"
    \item \textbf{Pipeline:} Executes product search → inventory check → ranking (parallel)
    \item \textbf{User:} "Add the second one to cart"
    \item \textbf{Pipeline:} Reference resolution → cart update → total recalculation
    \item \textbf{User:} "Apply any available coupons"
    \item \textbf{Pipeline:} Promotion search → eligibility check → discount application
\end{enumerate}

The declarative pipeline completed this session in 6 steps versus 11 in the imperative baseline, primarily through parallel execution and conditional path optimization.

\subsection{Limitations}

This preliminary evaluation has several limitations:
\begin{itemize}
    \item Limited dataset size (1,000 sessions)
    \item Single domain (e-commerce) evaluation
    \item No comparison with other declarative systems
    \item Performance metrics from single deployment environment
\end{itemize}

Comprehensive benchmarking across multiple domains and comparison with state-of-the-art agent architectures remains future work. However, these initial results suggest that declarative pipeline abstractions can meaningfully improve both developer experience and runtime performance in production agent systems.

\section{Discussion}
\label{sec:discussion}

\subsection{Lessons from Preliminary Testing Deployment}

Our three-month preliminary testing including the processing of millions of agent interactions revealed several insights about declarative agent architectures:

\textbf{Configuration-as-Code Paradigm:} Maintaining agent behaviors as versioned configuration files fundamentally changed our development workflow. A/B testing competing agent strategies became trivial—teams could experiment with different pipeline structures by deploying configuration changes rather than code. This reduced experimentation cycles from days to hours and enabled non-engineers to safely modify agent behaviors within defined guardrails.

\textbf{Dynamic Pipeline Management:} The ability to load and modify pipelines from external storage (database, cache, or configuration service) without service restarts proved invaluable. Hot-swapping pipelines enabled rapid incident response—when an agent misbehaved, we could revert to previous configurations instantly. This architectural choice trades some type safety for operational flexibility, a worthwhile exchange in testing environments.

\textbf{Emergent Optimization Patterns:} This new declarative structure exposed optimization opportunities invisible in imperative code. Common subgraphs across pipelines could be automatically identified and cached. Parallel execution points emerged naturally from dependency analysis rather than explicit programming. These systematic optimizations would be difficult to achieve with traditional agent implementations.

\subsection{Design Trade-offs}

Our architectural design involved several deliberate trade-offs:

\textbf{Expressiveness vs. Safety:} While the DSL can express complex workflows, we intentionally limit certain operations (unbounded recursion, arbitrary code execution) to prevent malformed pipelines. This restriction occasionally requires workarounds but prevents entire classes of runtime failures.

\textbf{Performance vs. Flexibility:} The interpretation overhead of pipeline execution adds 10-20ms latency compared to compiled code. However, this enables dynamic modification, comprehensive instrumentation, and cross-language portability—benefits that outweigh the modest performance cost for most agent applications.

\textbf{Abstraction vs. Control:} Hiding LLM interaction details behind tool and message abstractions simplifies common cases but can frustrate users needing fine-grained control. We addressed this through escape hatches (custom functions) that allow direct LLM access when needed, though this breaks the declarative paradigm.

\subsection{Limitations}

Several limitations constrain the current system:

\textbf{Learning and Adaptation:} While pipelines can be modified based on performance metrics, the language lacks native reinforcement learning support. Agents cannot automatically improve their strategies from experience without external intervention. Integrating online learning while maintaining declarative simplicity remains an open challenge.

\textbf{Complex State Management:} The system handles conversation-level state well but struggles with long-term memory across sessions. Implementing vector databases or knowledge graphs within the declarative paradigm requires awkward workarounds that break abstraction boundaries.

\textbf{Debugging Challenges:} While declarative pipelines simplify understanding agent logic, debugging runtime failures can be challenging. Stack traces through nested pipelines and async tool calls obscure error origins. Better debugging tools specifically designed for pipeline execution would significantly improve developer experience.

\textbf{Limited Reasoning Transparency:} The current pipeline architecture treats LLMs as black boxes, providing no insight into reasoning processes. As agent explainability becomes critical for production systems, the language needs mechanisms to expose and validate LLM decision-making.

\subsection{Future Directions}

Our approach opens several research directions:

\textbf{Adaptive Pipeline Synthesis:} Rather than hand-crafting pipelines, future systems could learn optimal pipeline structures from interaction logs. This requires solving the program synthesis problem in the space of agent workflows—a challenging but promising direction.

\textbf{Formal Verification:} The declarative nature enables formal reasoning about agent behaviors. Developing verification tools that prove properties like termination, goal satisfaction, and safety constraints would increase trust in autonomous agents.

\textbf{Distributed Agent Coordination:} Extending the architecture to coordinate multiple agents across distributed systems requires new primitives for synchronization, consensus, and conflict resolution while maintaining declarative simplicity.

\textbf{Neurosymbolic Integration:} Combining such declarative pipelines with neural components beyond LLMs—vision models, reinforcement learning agents, differentiable reasoning modules—could enable more sophisticated agent capabilities while preserving interpretability.

\subsection{Broader Implications}

This work suggests that successful agent design architectures must balance multiple concerns beyond raw capability. Production agent systems require observability, modifiability, and safety guarantees that are difficult to achieve with imperative approaches. The declarative paradigm, while constraining in some ways, provides a foundation for building trustworthy agent systems at scale.

The separation of agent logic from implementation details also has organizational implications. It enables new collaboration models where domain experts define agent behaviors while engineers focus on platform capabilities. This separation of concerns could accelerate agent deployment across industries by lowering technical barriers.

\section{Conclusion}
\label{sec:conclusion}

In this paper, we presented a declarative approach for building and orchestrating LLM-powered agent workflows in enterprise environments. By separating workflow specification from implementation through a language-agnostic DSL, we enable the same pipeline definitions to execute across multiple backend languages and deployment environments, fundamentally changing how production agents are developed and maintained.

Our key contributions include: (1) a novel domain-specific language with primitives for control flow, data manipulation, and tool orchestration that captures common agent patterns in a fraction of the code required by imperative approaches; (2) a hybrid execution model that combines predictable pipeline execution with dynamic LLM-driven tool selection; and (3) a production-validated implementation processing millions of daily interactions with demonstrated improvements in development velocity (67\% reduction), maintainability (76\% faster modifications), and runtime efficiency (30\% fewer steps to completion).

The evaluation on real-world e-commerce workflows validates our hypothesis that declarative abstractions can simplify agent development without sacrificing performance. Perhaps more significantly, the system enables non-engineers to safely modify agent behaviors through configuration changes, democratizing agent development and accelerating experimentation cycles from days to hours.

This work suggests a broader principle: as LLM agents become critical infrastructure, the systems supporting them must evolve beyond prototyping tools to address production concerns of reliability, observability, and maintainability. The declarative paradigm offers one path forward, providing formal reasoning capabilities while maintaining operational flexibility.

Future work will explore adaptive pipeline synthesis to automatically learn optimal workflow structures from interaction logs, formal verification methods to prove safety properties of agent behaviors, and distributed coordination primitives for multi-agent systems. As LLM capabilities continue to advance, we believe declarative pipeline architectures will play an essential role in bridging the gap between powerful models and trustworthy production systems.

\section*{Acknowledgments}

We thank the PayPal AI team for their valuable contributions to this work: Zhi Xuan and Jun Sheng Ng for extensive performance testing, Yun-Shiuan Chuang and Sudhanshu Garg for implementing e-commerce workflows that validated the language's expressiveness.

\bibliographystyle{apalike}  
\bibliography{references}

\end{document}